\documentclass[journal]{IEEEtran}
\usepackage{cite}

\ifCLASSINFOpdf
  \usepackage[pdftex]{graphicx}
  \DeclareGraphicsExtensions{.pdf,.jpeg,.png}
\else
\fi
\usepackage[font=footnotesize]{caption}
\usepackage{subcaption}

\usepackage{amsmath}
\usepackage{amssymb}
\usepackage{gensymb}

\usepackage{array}
\usepackage{booktabs}

\usepackage{algorithm}
\usepackage{algorithmic}

\usepackage{mathtools}

 \usepackage[caption=false,font=footnotesize]{subfig}

\usepackage{stfloats}
\begin{document}
%

\title{Two-Stage IQ Imbalance Estimation and Compensation for AFDM Systems}
%

%
%
%


\author{
Zhenfeng~Huang\IEEEauthorrefmark{1},
Yitong~Liu\IEEEauthorrefmark{1},
Yuping~Yan\IEEEauthorrefmark{2},
and~Hongwen~Yang\IEEEauthorrefmark{1}%
\thanks{Zhenfeng Huang, Yitong Liu, and Hongwen Yang are with the School of Information and Communication Engineering, Beijing University of Posts and Telecommunications, Beijing 100876, China 
(e-mail: bupt\_hzf@bupt.edu.cn; liuyitong@bupt.edu.cn; yanghong@bupt.edu.cn).}%
\thanks{Yuping Yan is with the China Satellite Network Group Co., Ltd, Beijing 100089, China 
(e-mail: yp\_jx001@sina.com).}%
}

\maketitle

\begin{abstract}
Affine frequency division multiplexing (AFDM) is an emerging chirp-based multicarrier waveform with strong diversity in doubly selective channels, but practical systems suffer from transmitter and receiver IQ imbalance, causing image interference and performance degradation. This paper proposes a two-stage IQ imbalance estimation and compensation method for AFDM systems. First, a preamble-assisted iterative algorithm estimates the time-invariant IQ imbalance parameters by exploiting their slowly time-varying nature. Then, a joint channel estimation and data detection scheme combines basis expansion model (BEM)-based channel estimation with an improved LMMSE detector for interference suppression. Simulations show rapid convergence and near-ideal BER performance.
\end{abstract}


%

\section{Introduction}
%
%
%
%

 

\IEEEPARstart{C}{urrently}, space-air-ground integrated wireless networks have emerged as a key paradigm for next-generation mobile communication systems, with low-Earth orbit satellite communications poised to become a crucial component of 6G networks \cite{xiao_space-air-ground_2024,zhang_space-air-ground_2025,cui_space-air-ground_2022}.
Recent studies have shown that Affine Frequency Division Multiplexing (AFDM) achieves superior diversity performance in doubly selective channels while requiring lower pilot overhead compared to Orthogonal Time Frequency Space (OTFS) modulation, making it particularly suitable for high-mobility scenarios such as LEO satellite communications \cite{yin_affine_2025,benzine_affine_2023,bemani_affine_2023,wang_afdm_2025}.
AFDM is a chirp-based multicarrier waveform that generalizes orthogonal frequency division multiplexing (OFDM) by multiplexing symbols in the discrete affine Fourier transform (DAFT) domain. With properly selected DAFT parameters $c_1$ and $c_2$, AFDM can separate different delay paths and achieve full diversity in doubly selective channels \cite{bemani_afdm_2021}.

Despite these advantages, the practical deployment of AFDM systems faces significant challenges due to hardware impairments. 
In direct-conversion transceivers, gain and phase mismatches between the in-phase (I) and quadrature (Q) branches cause IQ imbalance, which introduces image interference, distorts the received signal, and severely degrades performance in high-mobility and low-SNR scenarios such as LEO satellite communications \cite{neelam_channel_2021,sandell_estimation_2021}.


For AFDM systems, research on IQ imbalance remains relatively limited. 
Sui \textit{et al.} \cite{sui_mimo-afdm_2026} established a unified input--output model for MIMO-AFDM systems with multiple hardware impairments, including transmitter IQ imbalance, and showed that AFDM is more robust to hardware impairments than OFDM.
Gunasekara and Bedeer \cite{gunasekara_analysis_2025} derived the input--output relationship of AFDM systems with receiver IQ imbalance and proposed a real-valued LMMSE detector to compensate for receiver-side IQ imbalance under ideal channel estimation.
Liu \textit{et al.} \cite{liu_analysis_2026} derived the input--output relationship of AFDM systems under joint TX and RX IQ imbalance, established a theoretical upper bound on the average bit error probability as a function of the IQ imbalance parameters, and proposed a cascaded compensation scheme that sequentially compensates for Rx and Tx IQ imbalance.
Nevertheless, the impact of IQ imbalance on practical channel estimation for AFDM systems remains unexplored, and the estimation of IQ imbalance parameters in doubly selective channels has not yet been addressed.

This letter aims to develop an effective IQ imbalance estimation and compensation framework for AFDM systems in doubly selective channels. Unlike existing studies, this work investigates IQ imbalance in practical AFDM systems with BEM-based channel estimation in \cite{benzine_models_2026}  and LMMSE detection, and further develops dedicated algorithms for parameter estimation and data-phase compensation. 
The contributions of this letter are summarized as follows. 

\textbf{(i) \textit{Parameter Estimation:}} A preamble-assisted iterative algorithm is proposed to estimate IQ imbalance parameters through successive channel estimation, parameter estimation, and interference cancellation. 
\textbf{(ii) \textit{Compensation and Detection:} }A data-phase joint channel estimation and detection scheme is developed by combining BEM-based channel estimation with an enhanced LMMSE detector for image-interference suppression and symbol recovery. 
\textbf{(iii) \textit{Simulation Validation:} }Simulations demonstrate fast convergence and significant BER improvement, especially for higher-order modulations and moderate-to-high SNR regimes.



\section{System Model}


\subsection{Fundamentals of AFDM}

The basic principle of AFDM modulation is described as follows. Let $\mathbf{x} \in \mathbb{C}^{N \times 1}$ denote the data symbol vector in the DAFT domain, where $N$ is the number of subcarriers. After inverse DAFT (IDAFT), the time-domain signal is given by
\begin{equation}
s[n] = \sum_{m=0}^{N-1} x[m] e^{j2\pi \left( c_1n^2+c_2m^2+\frac{nm}{N} \right)}/\sqrt{N},
\end{equation}
which can be written in matrix form as
\begin{equation}
\mathbf{s} = \mathbf{A}^H \mathbf{x} = \mathbf{\Lambda}_{c_1}^H \mathbf{F}^H \mathbf{\Lambda}_{c_2}^H \mathbf{x},
\end{equation}
where $\mathbf{\Lambda}_c = \mathrm{diag}\left(e^{-j2\pi c n^2}, n=0,\dots,N-1\right)$, $\mathbf{F}$ is the $N$-point DFT matrix 
 and $\mathbf{A} = \mathbf{\Lambda}_{c_2} \mathbf{F} \mathbf{\Lambda}_{c_1}$ is the DAFT matrix.
 
Unlike OFDM, AFDM employs a chirp-periodic prefix (CPP) instead of a cyclic prefix (CP) to combat multipath propagation and eliminate inter-symbol interference (ISI)  \cite{bemani_affine_2023}.

We consider a doubly selective channel with $P$ propagation paths. The received signal is given by
\begin{equation}
r[n] = \sum_{l=0}^{\infty} s[n-l] g_n(l) + w[n],
\end{equation}
where $w[n] \sim \mathcal{CN}(0,\sigma_n^2)$ and $g_n(l) = \sum_{i=1}^{P} h_i e^{-j2\pi f_i n} \delta(l - l_i)$ is the time-varying impulse response at time $n$ and delay $l$. Here, $h_i$, $f_i$, and $l_i$ denote the complex gain, Doppler frequency, and integer delay of the $i$-th path, respectively. 

After passing through the doubly selective channel and demodulation, the matrix-form input--output relationship can be written as
\begin{equation}
\mathbf{y}=\mathbf{H}_{\mathrm{eff}}\mathbf{x}+\mathbf{w},
\end{equation}
where $\mathbf{w} \sim \mathcal{CN}(0,\sigma_n^2 \mathbf{I}_N)$ and the channel matrix is given by $\mathbf{H}_{\mathrm{eff}}=\sum_{i=1}^P{h_i\mathbf{A\Gamma }_{\mathrm{CPP}_i}\mathbf{\Delta }_{f_i}\mathbf{\Pi }^{l_i}\mathbf{A}^{\mathrm{H}}}$ \cite{benzine_affine_2023}.


\subsection{AFDM System Under Transceiver IQ Imbalance}

In practical direct-conversion transceivers, gain and phase mismatches between the I and Q branches introduce IQ imbalance, which can be modeled in the complex baseband domain as \cite{Razavi2012RF}
\begin{equation}
\tilde{s}(t) = \gamma s(t) + \eta s^*(t),
\end{equation}
where $\tilde{(\cdot)}$ denotes the IQ-impaired signal, $\gamma = (\cos \phi + j\epsilon \sin \phi)/\sqrt{1+\epsilon^2}$ and $\eta = (\epsilon \cos \phi - j\sin \phi)/\sqrt{1+\epsilon^2}$ denote the direct and image coefficients, respectively, with $\epsilon$ and $\phi$ being the amplitude and phase imbalance parameters. Moreover, $|\gamma|^2 + |\eta|^2 = 1$, and $\gamma \approx 1$, $\eta \approx 0$ for small IQ imbalance.

We now consider the AFDM system in the presence of IQ imbalance at both TX and RX. 
Let $\gamma_T, \eta_T$ and $\gamma_R, \eta_R$ denote the IQ imbalance parameters at the transmitter and receiver, respectively.
After AFDM modulation and transmitter IQ imbalance, the transmitted signal is
\begin{equation}
\mathbf{s}_{\mathrm{TXIQ}} = \gamma_T \mathbf{s} + \eta_T \mathbf{s}^* = \gamma_T \mathbf{A}^H \mathbf{x} + \eta_T \mathbf{A}^T \mathbf{x}^*.
\end{equation}
After passing through the doubly selective channel, the received signal becomes
\begin{equation}
\mathbf{r} = \mathbf{H}\mathbf{s}_{\mathrm{TXIQ}} + \mathbf{w}
= \gamma_T \mathbf{H}\mathbf{A}^H \mathbf{x} + \eta_T \mathbf{H}\mathbf{A}^T \mathbf{x}^* + \mathbf{w}.
\end{equation}
\begin{figure*}[!b]
\hrulefill
\vspace{0em}
\begin{equation}
\begin{split}
\mathbf{y} &= \gamma_R \mathbf{A}\mathbf{r} + \eta_R \mathbf{A}\mathbf{r}^* \\
&= \gamma_T \gamma_R \mathbf{A}\mathbf{H}\mathbf{A}^H \mathbf{x}
+ \eta_T \gamma_R \mathbf{A}\mathbf{H}\mathbf{A}^T \mathbf{x}^*
+ \gamma_R \mathbf{A}\mathbf{w} 
+ \gamma_T^* \eta_R \mathbf{A}\mathbf{H}^* \mathbf{A}^T \mathbf{x}^*
+ \eta_R \eta_T^* \mathbf{A}\mathbf{H}^* \mathbf{A}^H \mathbf{x}
+ \eta_R \mathbf{A}\mathbf{w}^* \\
&= 
\underbrace{k_1 \mathbf{H}_{\mathrm{eff}} \mathbf{x}}_{\text{desired signal}}
+
\underbrace{
k_2 \mathbf{H}_{\mathrm{eff}} \mathbf{A}\mathbf{A}^T \mathbf{x}^*
+ k_3 \mathbf{A}\mathbf{A}^T \mathbf{H}_{\mathrm{eff}}^* \mathbf{x}^*
+ k_4 \mathbf{A}\mathbf{A}^T \mathbf{H}_{\mathrm{eff}}^* 
(\mathbf{A}\mathbf{A}^T)^* \mathbf{x}
}_{\text{IQ imbalance interference}}
+
\underbrace{\mathbf{v}}_{\text{noise}} .
\end{split}
\label{eq:AFDM_IQmm_y_mtxform}
\end{equation}
\vspace{0em}
\end{figure*}
After receiver IQ imbalance and DAFT demodulation, the received signal in the DAFT domain is given by (\ref{eq:AFDM_IQmm_y_mtxform}) at the bottom of this page,
where $k_1=\gamma _{\mathrm{T}}\gamma _{\mathrm{R}}$, $k_2=\eta _{\mathrm{T}}\gamma _{\mathrm{R}}$, $k_3=\gamma _{\mathrm{T}}^{*}\eta _{\mathrm{R}}$ and  $k_4=\eta _{\mathrm{T}}^{*}\eta _{\mathrm{R}}$.
Let $\mathbf{k} = [k_1, k_2, k_3, k_4]^T$, which satisfies the normalization condition $\|\mathbf{k}\|^2 = 1$.
The noise vector is given by $\mathbf{v}=\gamma _{\mathrm{R}}\mathbf{Aw}+\eta _{\mathrm{R}}\mathbf{Aw}^*$, which follows a noncircular complex Gaussian distribution, i.e., $\mathbf{v} \sim \mathcal{CN}(\mathbf{0}, \mathbf{C}_v, \mathbf{P}_v)$, where $\mathbf{C}_v = \sigma_n^2 \mathbf{I}_N$ and $\mathbf{P}_{\mathbf{v}}=2\gamma _{\mathrm{R}}\eta _{\mathrm{R}}\sigma _{n}^{2}\mathbf{AA}^{\mathrm{T}}=2\left( k_1k_3+k_2k_4 \right) \sigma _{n}^{2}\mathbf{AA}^{\mathrm{T}}$.

It is worth noting that in (\ref{eq:AFDM_IQmm_y_mtxform}) the matrix $\mathbf{A}\mathbf{A}^T$ is dense, which severely destroys the sparsity of $\mathbf{H}_{\mathrm{eff}}$ \cite{gunasekara_analysis_2025}. As a result, significant interference is introduced among pilot and data symbols, which degrades the performance of channel estimation and data detection. Therefore, dedicated parameter estimation and compensation schemes are required.

Since the IQ imbalance parameters are determined by hardware and remain constant within an AFDM frame, a two-stage estimation and compensation strategy is adopted. In the first stage, the IQ imbalance parameter vector $\mathbf{k}$ is estimated using preamble symbols. In the second stage, the estimated $\mathbf{k}$ is used for channel estimation and data detection of subsequent data symbols.

\section{Preamble-Assisted IQ Imbalance Parameter Estimation}

Let $\mathbf{x}_p$ denote the preamble symbol vector in the DAFT domain, and let $\mathbf{y}_p$ denote the corresponding received vector after passing through the doubly selective channel and suffering from IQ imbalance.
Since existing channel estimation methods can effectively exploit the sparsity of $\mathbf{H}_{\mathrm{eff}}$ to reduce computational complexity, we first assume that IQ imbalance does not exist, i.e., $k_1^{(0)} = 1, \quad k_2^{(0)} = k_3^{(0)} = k_4^{(0)} = 0,$ and perform channel estimation directly to obtain an initial estimate $\mathbf{H}_{\mathrm{eff}}^{(0)}$. This initialization is reasonable because practical IQ imbalance is typically mild.

Equation (\ref{eq:AFDM_IQmm_y_mtxform}) can be rewritten as a linear model with respect to $\mathbf{k}$:
\begin{equation}
\mathbf{y}_p = \mathbf{R}^{(i)} \mathbf{k} + \mathbf{v},
\label{eq:preamble_y_p}
\end{equation}
where
\begin{align}
\mathbf{R}^{(i)} &=
\Bigl[
\mathbf{H}_{\mathrm{eff}}^{(i-1)} \mathbf{x}_p,\;
\mathbf{H}_{\mathrm{eff}}^{(i-1)} \mathbf{A}\mathbf{A}^{T} \mathbf{x}_p^*,\;
\mathbf{A}\mathbf{A}^{T} \!\left( \mathbf{H}_{\mathrm{eff}}^{(i-1)} \mathbf{x}_p \right)^*, \nonumber \\
&\qquad
\mathbf{A}\mathbf{A}^{T} \!\left( \mathbf{H}_{\mathrm{eff}}^{(i-1)} \mathbf{A}\mathbf{A}^{T} \right)^* \mathbf{x}_p
\Bigr].
\label{eq:R_i}
\end{align}

By applying real-valued decomposition, the above model in (\ref{eq:preamble_y_p}) can be expressed as
\begin{equation}
\begin{bmatrix}
\mathbf{y}_{p,\mathrm{R}} \\
\mathbf{y}_{p,\mathrm{I}}
\end{bmatrix}
=
\begin{bmatrix}
\mathrm{Re}\{\mathbf{R}^{(i)}\} & -\mathrm{Im}\{\mathbf{R}^{(i)}\} \\
\mathrm{Im}\{\mathbf{R}^{(i)}\} & \mathrm{Re}\{\mathbf{R}^{(i)}\}
\end{bmatrix}
\begin{bmatrix}
\mathbf{k}_{\mathrm{R}} \\
\mathbf{k}_{\mathrm{I}}
\end{bmatrix}
+
\begin{bmatrix}
\mathbf{v}_{\mathrm{R}} \\
\mathbf{v}_{\mathrm{I}}
\end{bmatrix}.
\label{eq:R_i_real}
\end{equation}
This can be compactly written as $\tilde{\mathbf{y}}_p = \tilde{\mathbf{R}}^{(i)} \tilde{\mathbf{k}} + \tilde{\mathbf{v}},
$ where $\tilde{\mathbf{v}} \sim \mathcal{N}(\mathbf{0}, \mathbf{C}_{\tilde{\mathbf{v}}}^{(i)})$, and
\begin{equation}
\mathbf{C}_{\tilde{\mathbf{v}}}^{(i)} =
\frac{1}{2}
\begin{bmatrix}
\mathrm{Re}\{\mathbf{C}_{\mathbf{v}} + \mathbf{P}_{\mathbf{v}}^{(i)}\} & -\mathrm{Im}\{\mathbf{C}_{\mathbf{v}} - \mathbf{P}_{\mathbf{v}}^{(i)}\} \\
\mathrm{Im}\{\mathbf{C}_{\mathbf{v}} + \mathbf{P}_{\mathbf{v}}^{(i)}\} & \mathrm{Re}\{\mathbf{C}_{\mathbf{v}} - \mathbf{P}_{\mathbf{v}}^{(i)}\}
\end{bmatrix},
\label{eq:C_v_i}
\end{equation}
where $\mathbf{P}_{\mathbf{v}}^{(i)} = 2\left( k_1^{(i)} k_3^{(i)} + k_2^{(i)} k_4^{(i)} \right) \sigma_n^2 \mathbf{A}\mathbf{A}^{T}.$

An LMMSE estimator is then applied to estimate $\tilde{\mathbf{k}}$:
\begin{equation}
\begin{aligned}
\hat{\tilde{\mathbf{k}}}^{(i)} &=
\tilde{\mathbf{k}}^{(0)} +
\left( \tilde{\mathbf{R}}^{(i)} \right)^{T}
\left(
\tilde{\mathbf{R}}^{(i)} \left( \tilde{\mathbf{R}}^{(i)} \right)^{T}
+ \mathbf{C}_{\tilde{\mathbf{v}}}^{(i)}
\right)^{-1} \\
&\qquad 
\left(
\tilde{\mathbf{y}}_p - \tilde{\mathbf{R}}^{(i)} \tilde{\mathbf{k}}^{(0)}
\right).
\end{aligned}
\label{eq:LMMSE_est}
\end{equation}
The complex-valued estimate $\hat{\mathbf{k}}^{(i)}$ is then reconstructed as $\hat{\mathbf{k}}^{(i)} =
\left[ \mathbf{I}_4 \;\; \mathbf{0}_4 \right] \hat{\tilde{\mathbf{k}}}^{(i)}
+ j \left[ \mathbf{0}_4 \;\; \mathbf{I}_4 \right] \hat{\tilde{\mathbf{k}}}^{(i)},$
followed by normalization $\mathbf{k}^{\left( i \right)}=\hat{\mathbf{k}}^{\left( i \right)}/\left\| \hat{\mathbf{k}}^{\left( i \right)} \right\| $.

After obtaining the estimate of $\mathbf{k}$, the interference terms in  (\ref{eq:AFDM_IQmm_y_mtxform}) can be mitigated as
\begin{align}
 &\mathbf{y}^{(i)}= (1-\rho)\mathbf{y}^{(i-1)} \nonumber \\
&\quad + \rho \Big(
\mathbf{y}^{(0)}
- k_2^{(i)} \mathbf{H}_{\mathrm{eff}}^{(i-1)} \mathbf{A}\mathbf{A}^{T} \mathbf{x}_p^* \nonumber
- k_3^{(i)} \mathbf{A}\mathbf{A}^{T} (\mathbf{H}_{\mathrm{eff}}^{(i-1)} \mathbf{x}_p)^* \\
&\qquad
- k_4^{(i)} \mathbf{A}\mathbf{A}^{T} (\mathbf{H}_{\mathrm{eff}}^{(i-1)} \mathbf{A}\mathbf{A}^{T})^* \mathbf{x}_p
\Big),
\label{eq:yp_update}
\end{align}
where $\rho$ is a damping factor.

Using the updated $\mathbf{y}^{(i)}$, a refined channel estimate $\mathbf{H}_{\mathrm{eff}}^{(i)}$ can be obtained. Through iterative channel estimation and interference cancellation, an accurate estimate of $\mathbf{k}$ can be achieved. The complete IQ imbalance parameter estimation procedure appears in Algorithm \ref{alg:iq_est}.

\begin{algorithm}[t]
\caption{Iterative Preamble-Assisted IQ Imbalance Parameter Estimation}
\label{alg:iq_est}
\begin{algorithmic}[1]
\REQUIRE  Preamble sequence $\mathbf{x}_p$, received preamble vector $\mathbf{y}_p$, noise power $\sigma_n^2$, tolerance $\varepsilon$, maximum iterations $N_{\mathrm{iter}}$
\ENSURE  IQ imbalance parameter vector $\mathbf{k}$
\STATE \textbf{Initialization:} $\mathbf{k}^{(0)} = [1,0,0,0]^T$, $\mathbf{y}^{(0)} = \mathbf{y}_p$
\FOR{$i = 1$ to $N_{\mathrm{iter}}$}
    \STATE \textbf{Channel estimation:} obtain $\mathbf{H}_{\mathrm{eff}}^{(i-1)}$ from $\mathbf{y}^{(i-1)}$
    \STATE \textbf{IQ imbalance parameter estimation:}
    \STATE \hspace{\algorithmicindent} \parbox[t]{\dimexpr\linewidth-\algorithmicindent}{Compute $\tilde{\mathbf{R}}^{(i)}$ using $\mathbf{x}_p$ and $\mathbf{H}_{\mathrm{eff}}^{(i-1)}$ according to \eqref{eq:R_i} and \eqref{eq:R_i_real}}
    \STATE \hspace{\algorithmicindent} \parbox[t]{\dimexpr\linewidth-\algorithmicindent}{Compute $\mathbf{P}_{\mathbf{v}}^{(i)}$ from $\mathbf{k}^{(i-1)}$}
    \STATE \hspace{\algorithmicindent} Compute $\mathbf{C}_{\tilde{\mathbf{v}}}^{(i)}$ according to \eqref{eq:C_v_i}
    \STATE \hspace{\algorithmicindent} \parbox[t]{\dimexpr\linewidth-\algorithmicindent}{Compute LMMSE estimate $\hat{\mathbf{k}}^{(i)}$ according to \eqref{eq:LMMSE_est}}
    \STATE \hspace{\algorithmicindent} $\mathbf{k}^{(i)} \gets \hat{\mathbf{k}}^{(i)} / \|\hat{\mathbf{k}}^{(i)}\|$
    \IF{$\|\mathbf{k}^{(i)} - \mathbf{k}^{(i-1)}\| < \varepsilon$}
        \STATE \textbf{break}
    \ENDIF
    \STATE \textbf{Interference cancellation:} compute $\mathbf{y}^{(i)}$ using \eqref{eq:yp_update}
\ENDFOR
\RETURN $\mathbf{k}$
\end{algorithmic}
\end{algorithm}

It should be noted that although channel estimation is performed using the preamble symbols, the obtained channel estimation result cannot be directly applied to subsequent data symbols. This is because the AFDM channel matrix cannot effectively distinguish different paths with identical delays but different Doppler shifts. Therefore, existing channel estimation methods typically estimate the entire channel matrix rather than individual path parameters \cite{yin_diagonally_2024}. As a result, the estimated channel cannot be directly extrapolated to future symbols.
Although the BEM approach based on discrete prolate spheroidal sequences (DPSS) in \cite{benzine_models_2026} allows channel prediction by extending basis functions, we observe that the extended DPSS basis exhibits severe oscillations, which significantly amplify the estimation error of BEM coefficients and render channel prediction unreliable.
Therefore, only the estimated IQ imbalance parameter vector $\mathbf{k}$ is utilized in subsequent channel estimation and data detection.

\section{IQ Imbalance Compensation}

Assume that within one AFDM frame, the symbols following the preamble are arranged according to the single-pilot scheme in \cite{bemani_affine_2023}, including pilot symbols, data symbols, and guard symbols. 
The data symbols are drawn from energy-normalized rectangular QAM constellations. 
Due to the presence of the dense matrix $\mathbf{A}\mathbf{A}^T$ in the interference terms of (\ref{eq:AFDM_IQmm_y_mtxform}), the pilot components in the received signal are affected by IQ imbalance interference. 

Following the idea of iterative interference cancellation, an initial channel estimate $\mathbf{H}_{\mathrm{eff}}^{(0)}$ is first obtained directly from the received signal.
Given the IQ imbalance parameters and the channel matrix, (\ref{eq:AFDM_IQmm_y_mtxform}) can be rewritten as a linear model with respect to $\mathbf{x}$. Define
\begin{equation}
\begin{aligned}
\mathbf{G}_1^{(i)} &= k_1 \mathbf{H}_{\mathrm{eff}}^{(i-1)} + k_4 \mathbf{A}\mathbf{A}^T \left( \mathbf{H}_{\mathrm{eff}}^{(i-1)} \mathbf{A}\mathbf{A}^T \right)^* \\
\mathbf{G}_2^{(i)} &= k_2 \mathbf{H}_{\mathrm{eff}}^{(i-1)} \mathbf{A}\mathbf{A}^T + k_3 \mathbf{A}\mathbf{A}^T \left( \mathbf{H}_{\mathrm{eff}}^{(i-1)} \right)^*,
\end{aligned}
\label{eq:G1_i_G2_i}
\end{equation}
then the real-valued equivalent model can be expressed as
\begin{equation}
\tilde{\mathbf{y}}^{(0)} = \tilde{\mathbf{G}}^{(i)} \tilde{\mathbf{x}} + \tilde{\mathbf{v}},
\label{eq:real_model}
\end{equation}
where
\begin{equation}
\tilde{\mathbf{G}}^{(i)} =
\begin{bmatrix}
\mathrm{Re}\{\mathbf{G}_1^{(i)} + \mathbf{G}_2^{(i)}\} & -\mathrm{Im}\{\mathbf{G}_1^{(i)} - \mathbf{G}_2^{(i)}\} \\
\mathrm{Im}\{\mathbf{G}_1^{(i)} + \mathbf{G}_2^{(i)}\} & \mathrm{Re}\{\mathbf{G}_1^{(i)} - \mathbf{G}_2^{(i)}\}
\end{bmatrix}.
\label{eq:G_real}
\end{equation}

An LMMSE estimator is then applied to estimate $\tilde{\mathbf{x}}$:
\begin{equation}
\begin{aligned}
\hat{\tilde{\mathbf{x}}}^{(i)} &=
\mathbf{\mu}_{\tilde{\mathbf{x}}}
+
\mathbf{C}_{\tilde{\mathbf{x}}}
\left( \tilde{\mathbf{G}}^{(i)} \right)^T
\left(
\tilde{\mathbf{G}}^{(i)} \mathbf{C}_{\tilde{\mathbf{x}}} \left( \tilde{\mathbf{G}}^{(i)} \right)^T
+ \mathbf{C}_{\tilde{\mathbf{v}}}
\right)^{-1} \\
&\qquad
\left(
\tilde{\mathbf{y}}^{(0)} - \tilde{\mathbf{G}}^{(i)} \mathbf{\mu}_{\tilde{\mathbf{x}}}
\right),
\label{eq:LMMSE_detect}
\end{aligned}
\end{equation}
where $\mathbf{\mu }_{\tilde{\mathbf{x}}}=\left[ \mathrm{Re}\left\{ \mathbf{\mu }_{\mathbf{x}}^{\mathrm{T}} \right\} ,\mathrm{Im}\left\{ \mathbf{\mu }_{\mathbf{x}}^{\mathrm{T}} \right\} \right] ^{\mathrm{T}},$ $\mathbf{\mu }_{\mathbf{x}}\in \mathbb{C} ^{N\times 1}$ whose entries are equal to pilot symbols at pilot positions and zero elsewhere. 
The covariance matrix is defined as $\mathbf{C}_{\tilde{\mathbf{x}}}=\frac{1}{2}\mathrm{blk}\mathrm{diag}\left\{ \mathbf{C}_{\mathbf{x}},\mathbf{C}_{\mathbf{x}} \right\} $ where $\mathbf{C}_{\mathbf{x}} \in \mathbb{R}^{N \times N}$ is a diagonal matrix whose diagonal elements are equal to 1 at data symbol positions and 0 otherwise.
After obtaining $\mathbf{x}^{(i)}$ from \eqref{eq:LMMSE_detect}, interference cancellation is performed to update the received signal similarly to (\ref{eq:yp_update}).

Using the updated $\mathbf{y}^{(i)}$, a refined channel estimate $\mathbf{H}_{\mathrm{eff}}^{(i)}$ can be obtained. By iteratively performing channel estimation, data detection via \eqref{eq:LMMSE_detect}, and interference cancellation via \eqref{eq:yp_update}, the effects of IQ imbalance can be effectively mitigated.
The complete joint channel estimation and data detection procedure under IQ imbalance is summarized in Algorithm \ref{alg:iq_comp}.

\begin{algorithm}[t]
\caption{Joint Channel Estimation and Data Detection Under IQ Imbalance}
\label{alg:iq_comp}
\begin{algorithmic}[1]
\REQUIRE IQ imbalance parameter vector $\mathbf{k}$, received signal $\mathbf{y}^{(0)}$, prior mean $\mathbf{\mu}_{\tilde{\mathbf{x}}}$, covariance $\mathbf{C}_{\tilde{\mathbf{x}}}$, noise power $\sigma_n^2$, maximum iterations $N_{\mathrm{iter}}$
\ENSURE Estimated channel $\mathbf{H}_{\mathrm{eff}}$, detected symbols $\mathbf{x}$
\STATE Compute $\mathbf{C}_{\tilde{\mathbf{v}}}$ from $\sigma_n^2$ and $\mathbf{k}$
\FOR{$i = 1$ to $N_{\mathrm{iter}}$}
    \STATE \textbf{Channel estimation:} obtain $\mathbf{H}_{\mathrm{eff}}^{(i-1)}$ from $\mathbf{y}^{(i-1)}$
    \STATE \textbf{Data detection:}
    \STATE \hspace{\algorithmicindent} Compute $\mathbf{G}_1^{(i)}$ and $\mathbf{G}_2^{(i)}$ using \eqref{eq:G1_i_G2_i}
    \STATE \hspace{\algorithmicindent} Compute $\tilde{\mathbf{G}}^{(i)}$ according to \eqref{eq:G_real}
    \STATE \hspace{\algorithmicindent} Estimate $\hat{\tilde{\mathbf{x}}}^{(i)}$ using \eqref{eq:LMMSE_detect} and recover $\mathbf{x}^{(i)}$ from $\hat{\tilde{\mathbf{x}}}^{(i)}$ 
    \IF{$\mathbf{x}^{(i)} = \mathbf{x}^{(i-1)}$}
        \STATE \textbf{break}
    \ENDIF
    \STATE \textbf{Interference cancellation:} update $\mathbf{y}^{(i)}$ using \eqref{eq:yp_update} and $\mathbf{k}$
\ENDFOR
\RETURN $\mathbf{H}_{\mathrm{eff}}, \mathbf{x}$
\end{algorithmic}
\end{algorithm}

\begin{figure*}[!t]
\centering

\begin{minipage}[t]{0.32\textwidth}
\centering
\includegraphics[width=\linewidth]{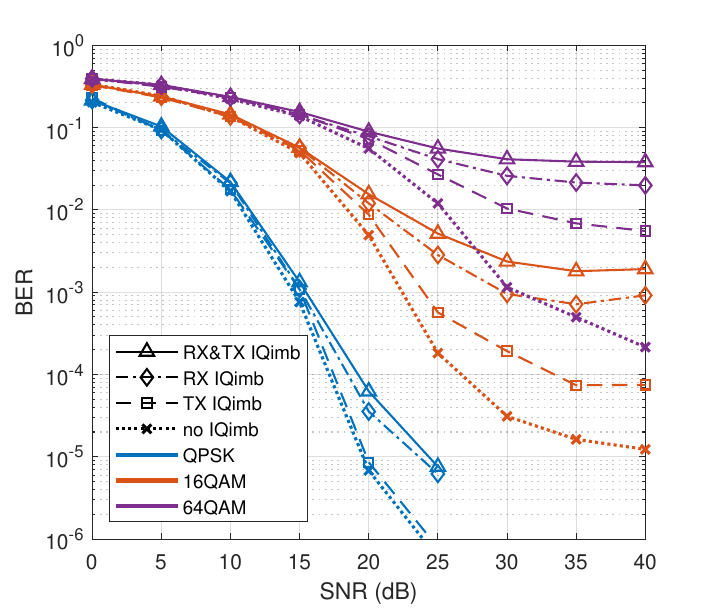}
\captionof{figure}{BER performance of AFDM under transmitter and receiver IQ imbalance.}
\label{fig:BER_IQI}
\end{minipage}
\hfill
\begin{minipage}[t]{0.32\textwidth}
\centering
\includegraphics[width=\linewidth]{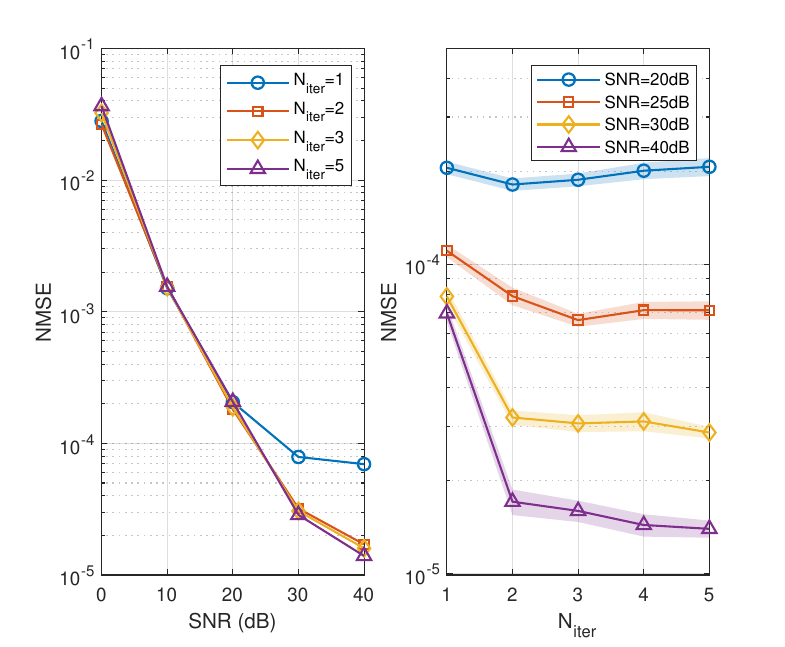}
\captionof{figure}{NMSE performance of the proposed IQ imbalance parameter estimation algorithm.}
\label{fig:NMSE_SNR_Niter}
\end{minipage}
\hfill
\begin{minipage}[t]{0.32\textwidth}
\centering
\includegraphics[width=\linewidth]{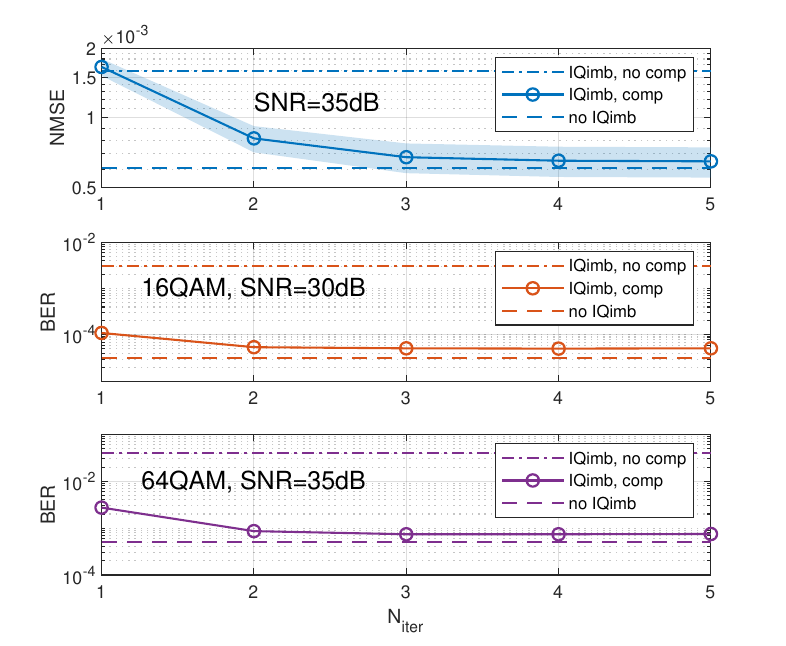}
\captionof{figure}{Performance of the proposed joint channel estimation and data detection algorithm.}
\label{fig:Joint_Performance}
\end{minipage}

\end{figure*}

\section{Simulation Results}

In this section, Monte Carlo simulations are conducted to evaluate the performance of the proposed IQ imbalance parameter estimation and compensation algorithms for AFDM systems. We consider an AFDM system with the number of subcarriers $N=256$, a sampling frequency of $10.24~\mathrm{MHz}$, and a carrier frequency of $f_c=4~\mathrm{GHz}$. The system is deployed on a low Earth orbit satellite at a height of  $h=500~\mathrm{km}$, where the maximum Doppler shift is $f_D=100~\mathrm{kHz}$.
The channel model follows the off-grid doubly sparse linear time-varying channel model in \cite{benzine_models_2026}, with maximum delay $l_{\max}=3$ and maximum integer Doppler $\alpha_{\max}=2$. In the delay-Doppler grid, $N_p=3$ grid points are randomly activated. Each activated grid point represents a cluster of paths sharing the same delay and integer Doppler but having different fractional Doppler components. Each cluster contains $N_D=3$ subpaths, resulting in a total number of paths $P=N_p N_D$. The complex gain of each path follows a zero-mean Gaussian distribution with variance $1/P$, and the fractional Doppler is uniformly distributed over $\left[-\frac{1}{2}, \frac{1}{2}\right]$. In the simulations, we set $\xi_{\nu}=3$, the AFDM parameter $c_1$ is determined according to $c_1 = \frac{2(\alpha_{\max} + \xi_\nu) + 1}{2N}$ and $c_2=\sqrt{2}/2$ \cite{benzine_affine_2023}.

Considering that several typical IQ modulators and demodulators provide IQ amplitude imbalance and phase imbalance parameters in their data sheets (as shown in Table \ref{tab:IQ_params}), we set the amplitude imbalance parameter $\epsilon=0.5~\mathrm{dB}$ and phase imbalance parameter $\phi=1.5^\circ$ as representative values to evaluate the BER performance under IQ imbalance. Unless otherwise specified, these parameters are used throughout the simulations.

\begin{table}[!t]
\centering
\caption{Typical IQ Imbalance Parameters of Practical Modulators and Demodulators}
\label{tab:IQ_params}
\begin{tabular}{cccc}\toprule

\textbf{Device} & $\epsilon$ (dB)& $\epsilon$ (\%)& $\phi$ (deg)\\\midrule
ADL5375-15 \cite{ADL5375}& 0.10 & 1.16 & 1.49 \\
ADMV4540 \cite{ADMV4540}& 0.50 & 5.93 & 1.6 \\
LTC5594 \cite{LTC5594}& 0.44 & 5.20 & 1.0 \\
MAX2022 \cite{max2202}& 0.30 & 3.51 & 0.5 \\ \bottomrule

\end{tabular}
\end{table}

Fig.~\ref{fig:BER_IQI} illustrates the BER performance of the AFDM system under transmitter and receiver IQ imbalance when the BEM-based channel estimation method in \cite{benzine_models_2026} and the LMMSE detector are employed. The results show that, under the same level of IQ imbalance, receiver IQ imbalance causes more severe BER degradation than transmitter IQ imbalance. Moreover, the performance loss is strongly dependent on the modulation order. For higher-order QAM schemes (e.g., 16QAM and 64QAM), the performance degradation induced by IQ imbalance becomes prohibitive.

Fig.~\ref{fig:NMSE_SNR_Niter} shows the NMSE of the estimated IQ imbalance parameters versus SNR and iteration number, averaged over $10^3$ independent channel realizations, with shaded regions denoting the $95\%$ confidence intervals. The proposed algorithm converges within a few iterations, and $3$ iterations provide a favorable performance--complexity trade-off. 
Fig.~\ref{fig:Joint_Performance} presents the channel estimation NMSE and BER of the proposed joint channel estimation and data detection algorithm, showing that its performance approaches the ideal hardware case after only a few iterations.

\begin{figure*}[!t]
\centering
\begin{minipage}[t]{0.30\textwidth}
\centering
\includegraphics[height=4.2cm,width=\linewidth,keepaspectratio]{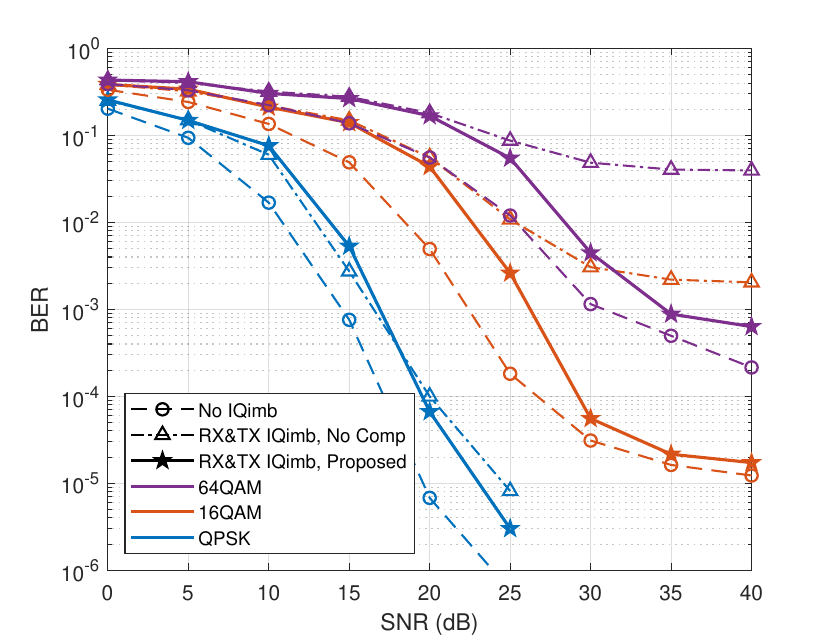}
\captionof{figure}{BER performance of the proposed IQ imbalance compensation algorithm versus SNR.}
\label{fig:BER_SNR_comp}
\end{minipage}%
\hspace{0.006\textwidth}%
\begin{minipage}[t]{0.66\textwidth}
\centering
\begin{minipage}[t]{0.49\linewidth}
\centering
\includegraphics[height=4.2cm,width=\linewidth,keepaspectratio]{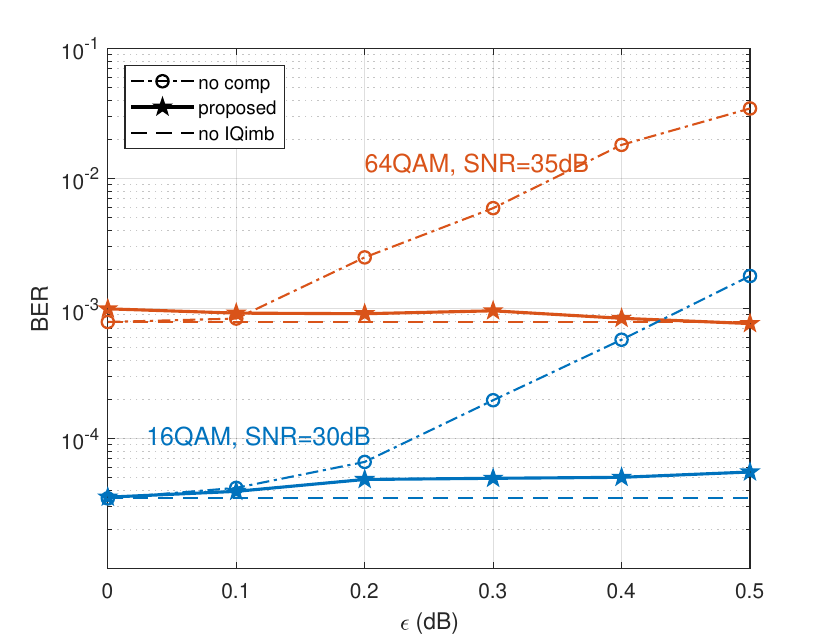}

\vspace{0.3em}
{\footnotesize (a) Amplitude imbalance}
\end{minipage}%
\hspace{0.01\linewidth}%
\begin{minipage}[t]{0.49\linewidth}
\centering
\includegraphics[height=4.2cm,width=\linewidth,keepaspectratio]{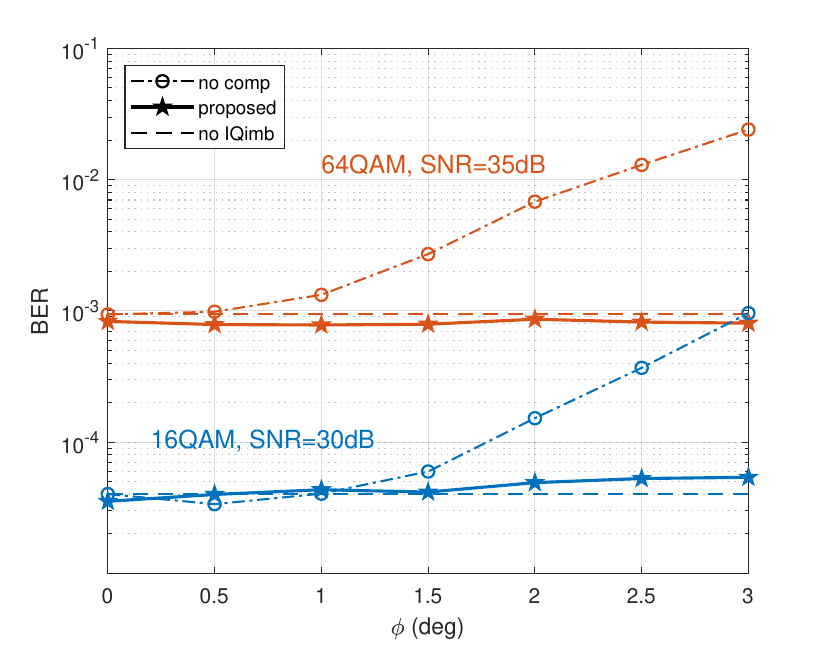}

\vspace{0.3em}
{\footnotesize (b) Phase imbalance}
\end{minipage}

\captionof{figure}{BER performance under different IQ imbalance severity.}
\label{fig:BER_IQI_severity}
\end{minipage}
\end{figure*}

In Fig.~\ref{fig:BER_SNR_comp}, the BER performance of the proposed IQ imbalance compensation algorithm is evaluated over different SNR values with $\epsilon=0.5~\mathrm{dB}$ and $\phi=1.5^\circ$. Fig.~\ref{fig:BER_IQI_severity} further investigates the BER performance under varying IQ imbalance levels by fixing $\phi=0$ or $\epsilon=0$ and varying the other parameter. The simulation results show that the proposed algorithm effectively mitigates the performance degradation caused by IQ imbalance at relatively high SNR. In particular, for higher-order modulations such as 16QAM and 64QAM, the compensated BER performance closely approaches that of the ideal hardware case. 



\section{Conclusion}


This letter proposed a two-stage IQ imbalance estimation and compensation framework for AFDM systems over doubly selective channels. In the first stage, a preamble-assisted iterative algorithm was developed to estimate the time-invariant IQ imbalance parameters through successive channel estimation, LMMSE-based parameter estimation, and interference cancellation. In the second stage, the estimated IQ parameters were incorporated into an iterative channel estimation and data detection scheme for data transmission, where an improved LMMSE detector was used to mitigate the interference term.
Extensive Monte Carlo simulations verified the effectiveness and robustness of the proposed scheme, showing that the proposed compensation algorithm achieves rapid convergence within approximately three iterations and maintains strong performance across a wide range of IQ imbalance levels.

\ifCLASSOPTIONcaptionsoff
  \newpage
\fi



%

\bibliographystyle{IEEEtran} 
\bibliography{bibtex/bib/references}




%








\end{document}